\documentclass[]{aastex631}

\usepackage{amsmath}

\begin{document}

\title{Formation Mechanism of Laser-Driven Magnetized ``Pillars of Creation"}

\correspondingauthor{Lifeng Wang, Bin Qiao}
\email{wang\_lifeng@iapcm.ac.cn, bqiao@pku.edu.cn}

\author{Zhu Lei}
\affiliation{Institute of Applied Physics and Computational Mathematics, Beijing 100094, China}

\author{Lifeng Wang$^*$}
\affiliation{Institute of Applied Physics and Computational Mathematics, Beijing 100094, China}
\affiliation{Center for Applied Physics and Technology, HEDPS, and State Key Laboratory of Nuclear Physics and Technology, School of Physics, Peking University, Beijing 100871, China}

\author{Jiwei Li}
\affiliation{Institute of Applied Physics and Computational Mathematics, Beijing 100094, China}
\affiliation{Center for Applied Physics and Technology, HEDPS, and State Key Laboratory of Nuclear Physics and Technology, School of Physics, Peking University, Beijing 100871, China}

\author{Shiyang Zou}
\affiliation{Institute of Applied Physics and Computational Mathematics, Beijing 100094, China}

\author{Junfeng Wu}
\affiliation{Institute of Applied Physics and Computational Mathematics, Beijing 100094, China}

\author{Zhonghai Zhao}
\affiliation{Center for Applied Physics and Technology, HEDPS, and State Key Laboratory of Nuclear Physics and Technology, School of Physics, Peking University, Beijing 100871, China}

\author{Wei Sun}
\affiliation{Department of Nuclear physics, China Institute of Atomic Energy, P.O. Box 275(7), Beijing 102413, China}

\author{Wenqiang Yuan}
\affiliation{Center for Applied Physics and Technology, HEDPS, and State Key Laboratory of Nuclear Physics and Technology, School of Physics, Peking University, Beijing 100871, China}

\author{Longxing Li}
\affiliation{Center for Applied Physics and Technology, HEDPS, and State Key Laboratory of Nuclear Physics and Technology, School of Physics, Peking University, Beijing 100871, China}

\author{Zheng Yan}
\affiliation{Institute of Applied Physics and Computational Mathematics, Beijing 100094, China}

\author{Jun Li}
\affiliation{Institute of Applied Physics and Computational Mathematics, Beijing 100094, China}

\author{Wenhua Ye}
\affiliation{Institute of Applied Physics and Computational Mathematics, Beijing 100094, China}

\author{Xiantu He}
\affiliation{Institute of Applied Physics and Computational Mathematics, Beijing 100094, China}
\affiliation{Center for Applied Physics and Technology, HEDPS, and State Key Laboratory of Nuclear Physics and Technology, School of Physics, Peking University, Beijing 100871, China}

\author{Bin Qiao$^*$}
\affiliation{Center for Applied Physics and Technology, HEDPS, and State Key Laboratory of Nuclear Physics and Technology, School of Physics, Peking University, Beijing 100871, China}
\affiliation{Frontiers Science Center for Nano-optoelectronic, Peking University, Beijing 100094, China}

\begin{abstract}
Pillars of Creation, one of the most recognized objects in the sky, are believed to be associated with the formation of young stars. However, so far, the formation and maintenance mechanism for the pillars are still not fully understood due to the complexity of the nonlinear radiation magneto-hydrodynamics (RMHD). \textcolor{black}{Here, under laboratory laser-driven conditions, we studied the self-consistent dynamics of pillar structures in magnetic fields by means of two-dimensional (2D) and three-dimensional (3D) RMHD simulations, and these results also support our new experimental scheme.} We find only when the magnetic pressure and ablation pressure are comparable, the magnetic field can significantly alter the plasma hydrodynamics. For medium magnetized cases ($\beta_{initial} \approx 3.5$), \textcolor{black}{the initial magnetic fields undergo compression and amplification. This amplification results in the magnetic pressure inside the pillar becoming large enough to support the sides of the pillar against radial collapse due to pressure from the surrounding hot plasma. This effect is particularly pronounced for the parallel component ($B_y$), which is consistent with observational results.} In contrast, a strong perpendicular ($B_x, B_z$) magnetic field ($\beta_{initial} < 1$) almost remains its initial distribution and significantly suppresses the expansion of blow-off gas plasma, leading to the inability to form pillar-like structures. The 3D simulations suggest that the bending at the head of `Column \uppercase\expandafter{\romannumeral1}' in pillars of creation may be due to the non-parallel magnetic fields. After similarity scaling transformation, our results can be applied to explain the formation and maintenance mechanism of the pillars, and can also provide useful information for future experimental designs.
\end{abstract}

\keywords{Laboratory astrophysics (2004), Plasma physics (2089), Magnetic fields (994), H II regions (694), Interstellar medium (847)}

\section{Introduction} \label{sec:intro}

\textcolor{black}{The interstellar medium is photo-ionized by ultraviolet (UV) radiation from nearby massive stars, as documented by \citet{oort1955acceleration} and \citet{pound2003looking}. Most of the UV radiation in this process comes from two stars of spectral type O5 V and O5.5 V in the nearby young cluster NGC 6611, as noted by \citet{pound1998molecular}. This photo-ionization is a widespread phenomenon in the universe, resulting in the formation of over-pressurized bubbles of ionized gas known as $\rm H~{\uppercase\expandafter{\romannumeral2}}$ regions \citep{Chur_2002}. Over a period of several million years, this process may give rise to the formation of thousands of stars due to the UV radiation ionization \citep{white1999eagle,kalari2018pillars,robertson2010early}.} The Pillars of Creation (POC, as shown in Fig.~\ref{setup}(b), also called `elephant trunks'), located in Eagle Nebula, are typical representative objects of $\rm H~{\uppercase\expandafter{\romannumeral2}}$ region and the reason why they are named so is that gas and dust are creating new stars, and they are also eroded by the UV radiation of the nearby stars. Recently, the James Webb Space Telescope captured a new high-resolution image of the POC, and the detail for the formation and development of young stars can be seen at the edges of these pillars \citep{webb_report_1,webb_report_2}. 

Although a lot of observation \citep{hester1996hubble,indebetouw2007embedded,pattle2018first} and numerical results \citep{lim20033d,mackey2010dynamical,mackey2011effects,hennebelle2019role} have been acquired, \textcolor{black}{the formation and maintenance mechanism of these pillars are }still not fully understood due to the nonlinear RMHD process. The prevailing view is that the primary physical process for the pillar formation is related to the ``rocket effect" \citep{spitzer1954behavior}, which is similar to the compression process of indirect-drive Inertial Confinement Fusion (ICF) fuel \citep{atzeni2004physics}. The UV radiation energy is absorbed near the cloud surface, which can lead to continued ablating of the surface and generating ionized blow-off flows away from the cloud surface. Because the ablation pressure generated by the ``rocket effect" is several orders of magnitude higher than the initial pressure, strong shocks are driven into the cloud, compressing and heating it. There exist about three models for the formation of these pillars, namely the instability model, cometary model, and shielding model \citep{remington2006experimental}. \textcolor{black}{The instability model suggested by Spitzer Jr. 1954 \citep{spitzer1954behavior} mainly pertains to the Rayleigh-Taylor instability (RTI) that continues to grow at the ablation front due to preexisting density fluctuations or fluctuations in UV radiation. The pillars are analogous to the ``spikes" of a heavy fluid penetrating through a light fluid, as noted by previous work \citep{frieman1954elephant}. This analogy is also applicable to the ablative RTI during the compression process of ICF \citep{ye2010spike,wang2014weakly,wang2017theoretical}.} The cometary model is that the pillars are formed by stand-alone dense clumps. A portion of the clumped material is pushed away from the radiation source by ablation pressure, forming a comet-like structure \citep{lefloch1994cometary}. The shielding model assumes that there are preexisting dense clumps within the low-density surrounding cloud, and the ablation pressure cannot push these dense clumps. These clumps roughly stay at their initial positions, while the rest of the cloud is moved a long distance by the ablation pressure, forming pillar structures \citep{williams2001hydrodynamics}. We can find that the latter two models are very similar, and the main difference is whether there is a background cloud. In recent years, the shielding model has received more attention, and a lot of numerical simulations are carried out based on this model \citep{lim20033d,pound2007pillars,mackey2011effects}.

Another confusing point about Eagle Nebula is the “missing stiffness” \citep{ryutov2005two,remington2006experimental}. According to the observation results \citep{levenson2000hot,ryutov2002scaling}, the material inside the clouds is very cold ($10- 30 ~  \mathrm{K}$), and the number density of the unshocked material is about $10^3 - 10^4 ~ \mathrm{cm^{-3}}$. However, the evaporated blow-off gas from the surface is very hot ($10^4 ~\mathrm{K}$), and the ablation pressure is about two orders of magnitude higher than the gas pressure inside the cloud, which causes the clouds or pillars to collapse to a higher density ($10^6 ~\mathrm{cm^{-3}}$), contrary to observations. Some models, such as the quasi-homogeneous magnetic field model or the magnetostatic turbulence model \citep{ryutov2005two}, have been proposed to solve this paradox, and magnetic fields are believed to play a key role in these models. The near-infrared extinction observations \citep{sugitani2007near} of the POC show magnetic fields are aligned along the pillars and the first high-resolution, submillimeter-wavelength polarimetric observations \citep{pattle2018first} also show that the magnetic field runs along the length of the pillars and the strength is on the order $100 ~ \mathrm{\mu Gs}$. These results suggest that the magnetic field is important in the formation of pillar-like structures and it also may be propping up the pillars of creation \citep{pattle2018first}. Due to the limitations of astronomical observational facilities and the complex background, the finer magnetic field structure still remains indistinct.

Over the past two decades, laboratory astrophysics has become an important tool in astrophysics research \citep{remington1999modeling,remington2006experimental,lebedev2019exploring}, and it can help us to investigate the individual specific factor on the formation of pillars under the conditions of scaling laws \citep{ryutov1999similarity,ryutov2001magnetohydrodynamic}. A novel experiment scheme was proposed to create a long-duration X-ray source to model the UV radiation produced by O stars \citep{casner2015long}. Then they use this X-ray source to simulate the shielding model in OMEGA and NIF facility \citep{kane2015dynamics,pound2017scaled}. The experiment results show a strong shock is launched into the CH foam, and the high-density clumps still roughly stay at the initial positions, forming a short pillar-like structure. 

However, the effect of magnetic fields on the formation of a pillar-like structure is still a subject of debate, and it also has not been studied in laboratory experiments. \textcolor{black}{In this paper, we use 2D and 3D RMHD simulations to study the detailed hydrodynamics process for the formation of the POC within the magnetic field. And these simulation results also provide information for future experimental designs.
To mimic the UV radiation in Eagle Nebula, a long-duration X-ray source can be achieved through an array of small laser-driven cavities \citep{casner2015long}. A low-density foam with preexisting spherical dense clumps is illuminated by the X-rays, and three materials (pure plastic (CH), doping aluminum (Al), and doping gold (Au)) foam targets are used in our RMHD simulations to study the radiation cooling effect.} The simulation results show that a strong shock can be launched in low-Z material (CH and Al), but the pillar-like structure collapses radially inward under the pressure of the surrounding hot plasma, unable to form a long collimated pillar structure. However, the existence of a magnetic field, especially the field parallel to the pillar structure ($B_y$ component in this work), can effectively support the sides of pillars against collapsing, which is consistent with the observation results \citep{sugitani2007near,pattle2018first}. The perpendicular components ($B_x, B_z$ in Fig.\ref{setup}(a)) of the magnetic field inside the foam are compressed and amplified to provide the balance with the ablation pressure. According to the 3D RMHD simulations, an asymmetrical pillar may be formed when the direction of the initial magnetic field is not perfectly parallel to the direction of shock propagation, which is similar to the `Column \uppercase\expandafter{\romannumeral1}' in the POC. These findings are confirmed through RMHD simulations and the similarities between laboratory and astrophysical systems \citep{ryutov2001magnetohydrodynamic} suggest that our results can be applied to explore the formation mechanism of pillars of creation.

\section{experimental scheme and simulation setup}

\begin{figure}[htb]
    \centering
    \includegraphics[width=16cm]{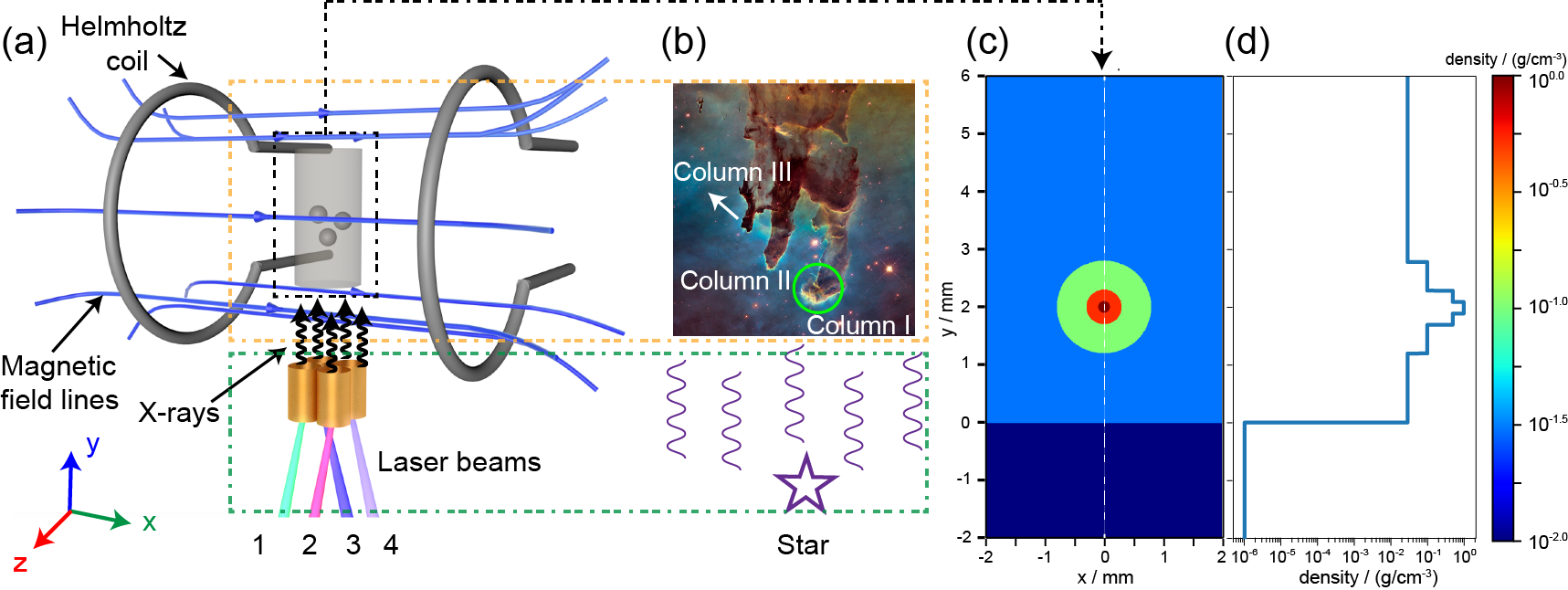} 
    \caption{{\bf \textcolor{black}{Schematic view of the experimental setup.}} A long-duration X-ray source is achieved through an array of small laser-driven cavities. These cavities are irradiated by a series of laser beams (as shown by ``1" to ``4" laser in panel (a)) in time (one after another), which can mimic the UV radiation from the nearby massive stars (as shown in panel (b)). The quasi-static magnetic field can be generated \textcolor{black}{through the pulse-power system (for weak magnetic fields) or laser-driven capacitor-coil target (for strong magnetic fields)}, and a uniform magnetic field is generated within the Helmholtz coil. We could change the direction of the coil to study the hydrodynamic evolution under different magnetic field components. The X-ray illuminates a low-density foam with preexisting spherical dense clumps, and three materials (CH foam, \textcolor{black}{doping Al foam, doping Au foam}) are used in our radiation simulation, which can be used to mimic the formation mechanism of the POC. The corresponding astrophysical system is shown in panel (b). Image credits: NASA, ESA, and the Hubble Heritage Team (STScI/AURA). \textcolor{black}{The initial density perturbation distribution is plotted in panel (c), and the colormap is corresponding to the logarithmic mass density in $\rm g/cm^3$. Panel (d) plots the initial radial distribution of foam density at x=0 (the position at the white dot line in panel (c))} }
    \label{setup}
\end{figure}

\textcolor{black}{As shown in Fig. \ref{setup}(a), We propose a new experimental scheme to study the formation of pillar structure within magnetic fields. The quasi-static magnetic field can be generated using either a pulse-power system (for weak magnetic fields) or a laser-driven capacitor-coil target (for strong magnetic fields). A uniform magnetic field is produced within the Helmholtz coil. And the hydrodynamic evolution under different magnetic field components can be achieved by changing the direction of the coil. When simulating the formation of pillars in the Eagle Nebula, it is more appropriate for high-power laser facilities to illuminate the target with X-rays from hohlraums rather than directly irradiating the target with lasers. Direct laser irradiation would introduce complications that do not exist in astrophysical systems, such as laser-plasma interaction instability and non-local electron heat transport \citep{kruer2000interaction, marocchino2013comparison}. A long-duration X-ray source is achieved through an array of small laser-driven cavities. These cavities are irradiated by a series of laser beams in time (``1" to ``4, one after another), and the radiation temperature ($T_r$) of X-rays is about $60 - 100 ~ \mathrm{eV}$.} Such a platform can be used to mimic the formation of the pillars in the Eagle Nebula and the corresponding astrophysical system is shown in Fig.~\ref{setup}(b), and the similarity scaling laws between the astrophysical systems and laboratory systems are given in Section \ref{sec:discussion}.

\textcolor{black}{Based on such experimental setup, we conducted numerical radiation simulations.} The 2D and 3D RMHD simulations are carried out by using the FLASH code \citep{fryxell2000flash}, which includes many high-energy-density physics modeling capabilities, such as laser energy deposition, multi-temperature ($ T_e \neq T_i \neq T_{rad} $), electron thermal conduction, and radiation transport, etc. In 2D RMHD simulations, the simulation domain size is set as $\rm (x,y) = 4000 ~ \mu m \times 8000 ~ \mu m $ with adaptive mesh refinement applied and the highest resolution $\rm 8 \mu m $ achieved. \textcolor{black}{And the simulation geometry is Cartesian geometry.} For 3D RMHD simulations, the simulation domain size is set as $\rm (x,y,z) = 4000 ~\mu m \times 8000 ~\mu m \times 4000 ~\mu m$ with the highest resolution $\rm 40 ~\mu m $ achieved. The X-ray radiation with $T_r = 60 ~\mathrm{eV}$ is set at the bottom boundary to simulate the long-duration X-ray source by an array of small laser-driven cavities, as shown in Fig.~\ref{radiation-material}(g). The Courant-Friedrichs-Lewy (CFL) is set as 0.4 and a third-order interpolation is used to reach a balance between accuracy and stability in simulation. The equation of state (EoS) and opacity of different materials (CH, Al, Au) are calculated from the code BADGER \citep{heltemes2012badger} and IONMIX \citep{macfarlane1989ionmix} respectively. \textcolor{black}{For foam targets doped with Al or Au elements, our targets will comprise 70\% CH and 30\% high-Z metal element doping. Such the composition of the target is expected to significantly decrease the electrical conductivity\citep{alhamidi2022conductive}, and the magnetic fields generated by the coils can diffuse smoothly throughout the entire foam target.} 

The initial density of foam target is set as $\rho_{\scalebox{0.6}{\textit{foam}}}$$ = 30 ~ \rm mg/cm^{3}$. For the preexisting spherical dense clumps, \textcolor{black}{considering the distribution of target manufacturing in the experiment and astrophysical perturbations\citep{mackey2011effects}, we use a triple-density sphere as a perturbation in the simulation. The initial density perturbation distribution is plotted in Fig.\ref{setup}(c). The distribution consists of three distinct regions. The first region, located within a radius of $100 ~\mathrm{\mu m}$, is a highly dense core with a density of $\rho_{\scalebox{0.6}{\textit{core}}} = 1.05~ \rm g/cm^{3}$. The second region, with a radius ranging from $100 ~\mathrm{\mu m}$ to $300 ~\mathrm{\mu m}$, is the middle part of the target where the density is set to $0.5 \rm~ g/cm^{3}$. The third region, with a radius ranging from $300 ~\mathrm{\mu m}$ to $800 ~\mathrm{\mu m}$, is the outer part of the target where the density is set to $0.1 \rm ~ g/cm^{3}$. And the initial radial distribution for the density of the target at $x=0$ is plotted in Fig.\ref{setup}(d)} Due to the configuration requirement of the code ``FLASH'', a low-density ($\rm 1 \ \times \ 10^{-6} \ g/cm^3$) Helium (He) background plasma is set in other places in simulations. All the initial temperatures are set to be uniform as room temperature $290 ~ \rm K$. The boundary conditions for fluid and radiation transport are all set to open.

\section{simulation results}
\subsection{Radiation cooling for different materials}

\begin{figure}[htb]
    \centering
    \includegraphics[width=16cm]{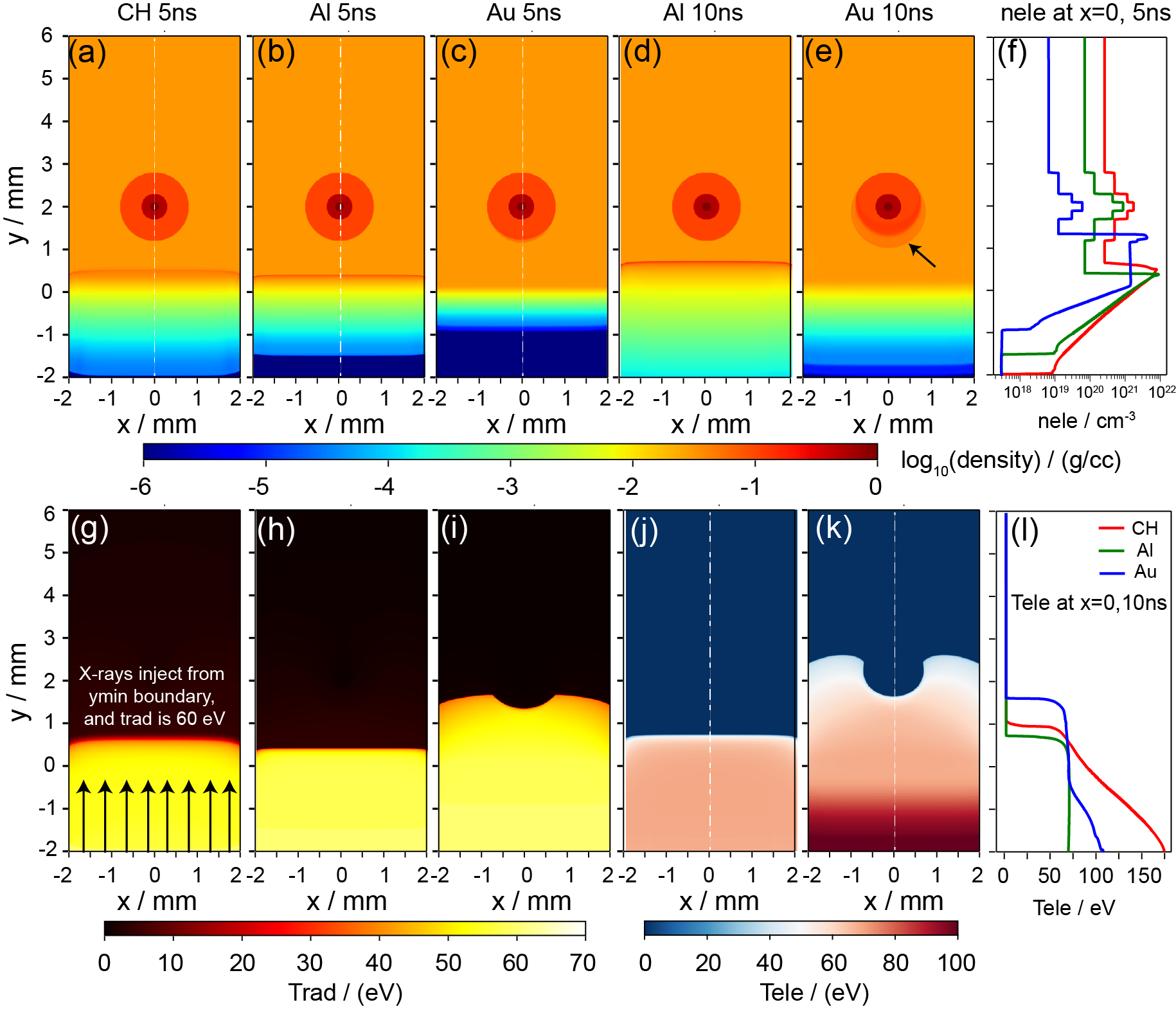} 
    \caption{{\bf \textcolor{black}{Radiation cooling for different materials.}} Panel (a) to (e), these color maps correspond to the logarithmic mass density in $\rm g/cm^3$ for CH, Al, and Au at different times. The electron number densities at $x=0$ (the position at white dot line in (a) to (c)) and $t = 5 ~\mathrm{ns}$ for different materials are shown in panel (f). Panel (g) to (i) are the corresponding radiation temperature distribution, and panels (j), (k) are the electron temperature distribution. Panel (l) plots the electron temperature at $x=0$ (the position at white dot line in (j) and (k)) and $t = 10 ~\mathrm{ns}$ of different materials.}
    \label{radiation-material}
\end{figure}

\textcolor{black}{In this section, we investigate the radiation-cooling effect on plasma dynamics by examining three materials (CH, Al, Au). According to previous work \cite{neufeld1995thermal}, the radiation cooling time, which is the ratio of thermal energy density to radiation power per unit volume, ranges from $10^2 - 10^3 ~\mathrm{years}$  for the POC. As for the hydrodynamic motion of POC, it has a characteristic time scale of $10^4 - 10^5 ~\mathrm{years}$, indicating that the shocked material cools down quickly due to radiation cooling. In laboratory experiments and simulations, we also need to consider the radiation-cooling effect for the formation mechanism of pillar structure. As mentioned in previous work \citep{tikhonchuk2008laboratory}, the radiative cooling time is inversely proportional to the nuclear charge Z when the temperature and number density are approximately equal. To investigate this effect, we selected three materials: pure CH plastic foam, Al-doped foam, and Au-doped foam. These materials represent low-Z, middle-Z, and high-Z materials, respectively.}

As shown in Fig.~\ref{radiation-material}(a), a high-density clump with a radius of $800 ~\mathrm{\mu m}$ is located at $y = 2 ~\mathrm{mm}$. X-rays inject from the bottom boundary of the simulation box and X-ray radiation is absorbed in the low-density foam surface by the photo-ionization process (Fig.~\ref{radiation-material}(a) to \ref{radiation-material}(e)). The ionized blow-off gas plasma begins to expand away from the foam surface, but the expansion velocities are different for the three materials. The CH gas plasma expands fastest, and the velocity is about $-4.7\times 10^7 ~\mathrm{cm/s}$. The expansion velocities of Al and Au gas plasma are $-3.0\times 10^7 ~\mathrm{cm/s}$ and $-6 \times 10^6 ~\mathrm{cm/s}$ respectively. Then the pressure from the ``rocket effect" compresses the foam, and according to the rocket model, the compression velocity is proportional to the mass ablation rate and blow-off gas velocity \citep{atzeni2004physics}. We can find that the compression velocity is fastest in CH plasma, and a strong shock is launched into the foam for CH and Al material, compressing and heating the foam. For the Au material, there is no shock formation within the foam due to strong radiation and low-speed blow-off gas.

Although there is no shock formation within the Au material, the preexisting dense clump within the foam is also ablated by X-rays, as indicated by the black arrow in Fig.~\ref{radiation-material}(e). According to the distribution of the radiation temperature, we find there is a significant difference in the deposition mode of X-ray energy between Au and CH (or Al) material. For CH or Al material, the X-ray energy is deposited in a very thin layer near the surface, and the temperature of undisturbed foam is much lower than the ablated gas plasma. While for the Au material, due to the strong radiation characteristics of Au materials, the energy of X-ray has been significantly deposited to the inner side, and the high-density clump is also preheated. However, as the density increases, X-rays cannot continue to deposit energy inward. As shown in Fig.~\ref{radiation-material}(i), the X-ray does not deposit energy into the interior of the high-density clump. This can be explained by the photon mean free path of radiation. The typical absorption coefficient \citep{johnston1973correct} in plasma can be expressed as $K=7.8 \times 10^{-9} \bar{Z} N_e^2 \ln \Lambda / [\nu^2\left(k_B T\right)^{3 / 2} \left(1-\nu_p^2 / \nu^2\right)^{1 / 2}]$, where $\ln \Lambda$ is Coulomb logarithm, $k_B T \approx 80 ~\mathrm{eV}$ are the Au plasma temperature (in \ref{radiation-material}(l)), $\bar{Z} \approx 25, ~ N_e$ are the average ionization and electron number density, $\nu$ and $\nu_p$ are radiation frequency and plasma frequency. The estimated radiation photon (energy $= 60 ~\mathrm{eV}$) mean free path $l_{free} \approx 1 / K \propto 1 / N_e^2$. As shown in \ref{radiation-material}(f), the electron number densities for the foam and clump of Au are $10^{20}~ \mathrm{cm^{-3}}$ and $4\times10^{21} ~\mathrm{cm^{-3}}$ respectively, and the corresponding radiation photon mean free paths are $7.7 ~ \mathrm{cm}$ and $50 ~\mathrm{\mu m}$. According to the above results, for strong radiation materials, a relatively higher-density foam is required to allow the X-ray energy to be deposited in a thin layer, then a shock can be launched into the foam, but it may need more time for the shock traversing the target.

As mentioned above, radiation cooling is significant to the evolution of the POC. For the laboratory system, the radiative cooling time is the ratio of the plasma thermal energy and radiated power. For simplicity, here we can ignore the line emission and account for the bremsstrahlung emission only \citep{tikhonchuk2008laboratory}. The bremsstrahlung emission power per unit energy and volume is scaled as $J_{br}(\nu, T) \propto T^{-\frac{1}{2}} n_e n_i Z^2 e^{-h\nu / kT} \ \mathrm{(erg  s^{-1}  cm^{-3}  Hz^{-1})}$ \citep{halverson1972bremsstrahlung}, where $\nu, T$ represent the photon frequency and temperature; and $n_e, n_i, Z$ are the number density of electron, ion, and the average ionic charge. We can integrate emission over the whole spectrum to get the power per volume scales as $P_{br} \propto T^{1/2} n_e n_i Z^2 \approx 1.7\times 10^{-32} Z n_e^2 T^{1/2} ~ \mathrm{W/cm^3} $ \citep{richardson20192019}. The energy density of plasma can be defined as \citep{nicolai2006plasma} $E_p = \frac{3}{2} n_e T \approx 2.4\times 10^{-19} n_e T ~ \mathrm{J/cm^3}$. Then the radiation cooling time can be calculated as $\tau_{rad} = E_p / P_{br} = 1.4\times 10^{22} T^{1/2} / (Zn_e) ~\mathrm{ns}$. Here, we mainly focus on the CH and Al material, where a strong shock is launched in these materials, and the related parameters are $T_{CH} = 100~\mathrm{ev},~ \bar{Z}_{CH} = 2.5, ~ n_{e, CH}\approx 6\times 10^{21} ~\mathrm{cm^{-3}}$, $T_{Al} = 70~\mathrm{ev},~ \bar{Z}_{Al} = 10, ~ n_{e, Al}\approx 6\times 10^{21} ~\mathrm{cm^{-3}}$ (see Fig.~\ref{radiation-material}(f) and \ref{radiation-material}(l)). Then the typical radiation cooling times can be obtained as $\tau_{rad, \scalebox{0.6}{\textit{CH}}} \approx 9.3 ~\mathrm{ns}, ~ \tau_{rad,\scalebox{0.6}{\textit{Al}}} \approx 1.9 ~\mathrm{ns}$ respectively. The hydrodynamic time is about $\tau_{\scalebox{0.6}{\textit{hd}}} = 120 ~\mathrm{ns}$ for the shock wave traversing all foam, and the radiation cooling parameter is $\chi = \tau_{rad} / \tau_{hd}$. For the POC, the radiation cooling parameter is in the range $\sim 10^{-3} - 10^{-2}$, and the radiation cooling parameter $\chi_{Al} \approx 1.5\times10^{-2}$ of Al material is closer to the astrophysical system, so for the experiment, CH foam doped with Al is a better choice. In the following section, we mainly focus on the Al material.

\subsection{The formation of pillar-like structure}

\begin{figure}[htb]
    \centering
    \includegraphics[width=16cm]{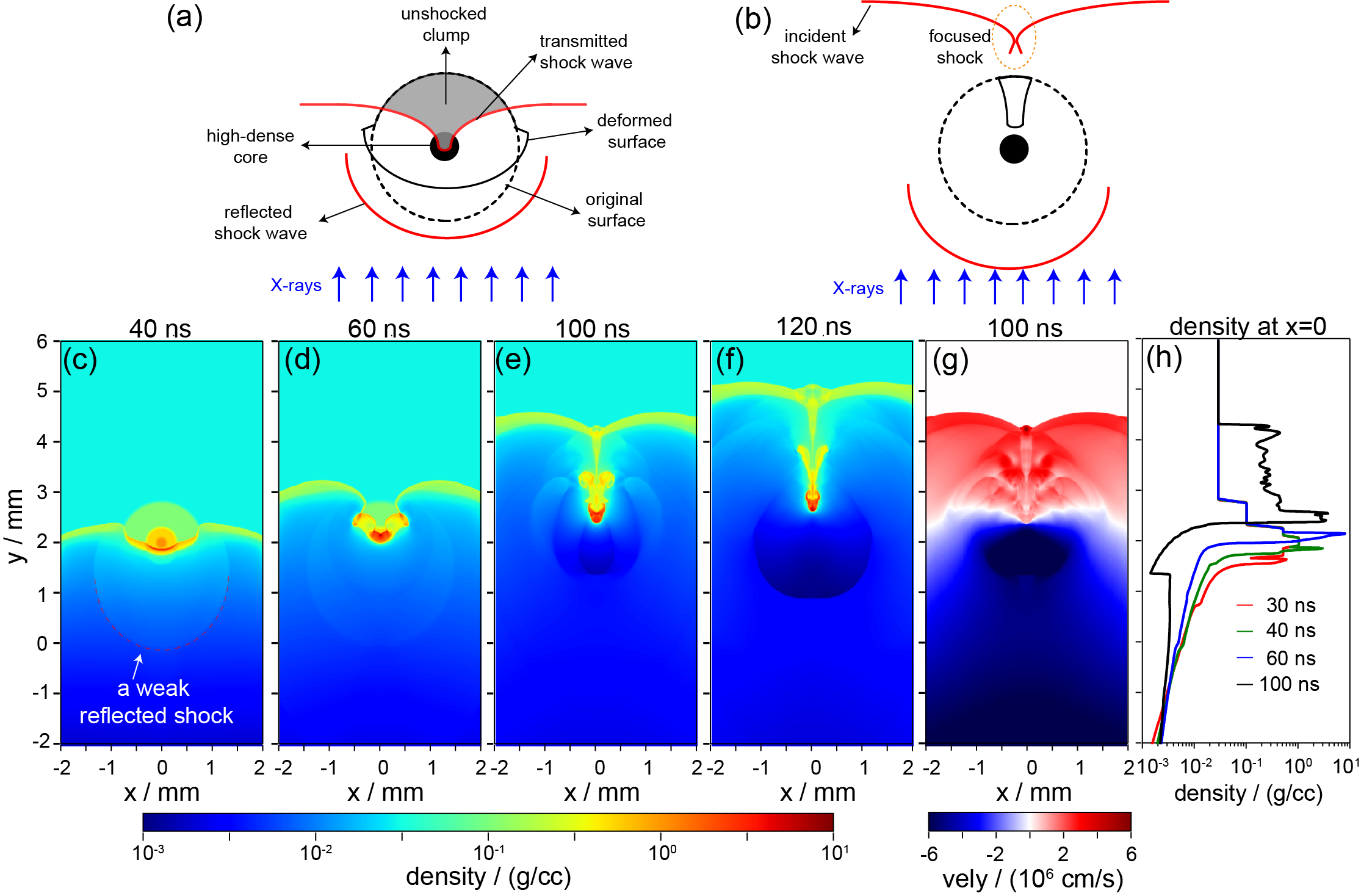} 
    \caption{{\bf \textcolor{black}{ The formation of pillar-like structure without magnetic fields.}} Panel (a) and (b) are the schematic views of the interaction between shock and clump, (a) during initial shock wave transit, and (b) after initial shock wave transit. Panel (c) to (f), these color maps correspond to the logarithmic mass density in $\rm g/cm^3$ for Al at 40 ns, 60 ns, 100 ns, and 120 ns respectively, and an apparent rarefaction wave travels back to the front surface after the shock has traversed the high-density clumps in panel (f). The distribution of velocity in the y-axis is shown in (g). The mass densities at $x=0$ for different times are shown in panel (h).}
    \label{pillar-form}
\end{figure}

The formation of the pillar-like structure is mainly due to the interaction of the shock wave with the pre-existing high-density clumps, which are similar to the shock–bubble interaction in fluid mechanism \citep{ranjan2011shock,niederhaus2008computational}. The Atwood number is defined as $A = (\rho_{\scalebox{0.6}{\textit{clump}}} - \rho_{\scalebox{0.6}{\textit{foam}}}) / (\rho_{\scalebox{0.6}{\textit{clump}}} + \rho_{\scalebox{0.6}{\textit{foam}}})$, where $\rho_{\scalebox{0.6}{\textit{clump}}},~ \rho_{\scalebox{0.6}{\textit{foam}}}$ represent the density of unshocked clumps and foam, and here we can obtain $A>0$. The specific heats $\gamma_{\scalebox{0.6}{\textit{clump}}}, ~\gamma_{\scalebox{0.6}{\textit{foam}}}$ are the same due to the same material, then we can get the sound speed in unshocked clumps to be smaller than foam ($c_{s,clump} < c_{s,foam}$). Thus the incident shock wave is refracted while crossing the curved upstream clump surface, owing to the change in sound speeds, and here the refraction is convergent so that the transmitted shock wave is concave curvature. As shown in Fig.~\ref{pillar-form}(a), an incident shock moves inward and the clump is compressed by this shock (compression phase). And according to the relationship: $R_0^{tt} = \gamma_{\scalebox{0.6}{\textit{foam}}}(\gamma_{\scalebox{0.6}{\textit{foam}}} + 1)M_i^2 / [\gamma_{\scalebox{0.6}{\textit{clump}}} - \gamma_{\scalebox{0.6}{\textit{foam}}} + \gamma_{\scalebox{0.6}{\textit{foam}}}(\gamma_{\scalebox{0.6}{\textit{clump}}}+1)M_i^2]$, where $M_i$ is the Mach number of the incident shock wave, we can obtain that $R_0^{tt} = 1$. The initial density ratio $R_0 = \rho_{\scalebox{0.6}{\textit{clump}}} / \rho_{\scalebox{0.6}{\textit{foam}}} > R_0^{tt} = 1$, thus the reflected wave is a shock wave \citep{niederhaus2008computational}. Fig.~\ref{pillar-form}(c) and \ref{pillar-form}(d) are corresponding to this compression phase. The shock structure is also similar to Fig.~\ref{pillar-form}(a), and a weak reflected shock wave moves back-forward to the shocked low-density foam.

When the incident shock pass through all clump, due to the high Atwood number ($A_{max} = (1.05-0.03)/(1.05+0.03) \approx 0.94$), portions of the shock front sweeping around the clump periphery are diffracted \citep{niederhaus2008computational}. They turn towards the axis so that the discontinuous surface is almost perpendicular to the interface. These diffracted shock waves may then converge with the transmitted shock wave at the downstream pole, leading to shock focusing, as shown in Fig.~\ref{pillar-form}(b). After this shock passes through the clump, a rarefaction wave travels back to the shocked materials, and then the whole clump is accelerated (acceleration phase). Figure \ref{pillar-form}(e) is corresponding to the acceleration phase. The shock passes through all clumps and is focused towards the axis, and a short pillar-like structure is formed after the high-density clump core, which is similar to the previous experiment result \citep{kane2015dynamics}. The velocity in the y-axis near the head (where the position of the high-density clump core) of this pillar is about $5\times10^5 ~\mathrm{cm/s}$, while at the rare surface of the pillar, the velocity is as high as $6\times10^6 ~\mathrm{cm/s} $. So this heavy clump core can be viewed as nearly stationary. In Fig. \ref{pillar-form}(g), we find the pillar-like structure collapses radially inward due to the surrounding hot plasma, which leads the pillar heads to be disconnected from the tails. Figure \ref{pillar-form}(h) plots the mass densities at $x=0$ for $t = 0,~30,~40,~60,~100~ \mathrm{ns}$ respectively, and both compression and acceleration phase are shown. In the compression phase, the maximum density at the shock position is about $3.5- 4$ times higher than the initial density, which means a strong shock is launched into the clump. 

The aspect ratio of length to width for the laboratory pillar is much smaller than the POC in Eagle Nebula. If there is no mechanism to support the sides of pillars against collapsing radially due to the surrounding hot plasma, the pillar may be destroyed in the end. The formation of the POC is related to many factors, such as multiple pre-existing disturbances \citep{lim20033d,pound2007pillars}, magnetic fields \citep{mackey2011effects,pattle2018first}, etc. In the following sections, we mainly focus on the effect of different components of magnetic fields.

\subsection{The effect of magnetic field}

\begin{deluxetable}{ccccccc}

    \tablecaption{\textcolor{black}{Different cases for the effect of magnetic field}}
    \label{table-case}
    \tablenum{1}
    
    \tablehead{\colhead{case types} & \colhead{case number} & \colhead{$B_x ~/~ (10^4 ~ \mathrm{Gs})$} & \colhead{$B_y ~/~(10^4 ~ \mathrm{Gs})$} & \colhead{$B_z ~/~ (10^4 ~ \mathrm{Gs})$} & \colhead{$\beta_{initial}$} & \colhead{dimension}} 
    
    \startdata
    &1       &20         &0          &0          &188    &2D        \\
    weakly magnetized &2       &0          &20         &0          &188    &2D        \\
    &3       &0          &0          &20         &188    &2D        \\
    \hline
    &4       &150        &0          &0          &3.35   &2D       \\
    medium magnetized&5       &0          &150        &0          &3.35   &2D       \\
    &6       &0          &0          &150        &3.35   &2D       \\
    \hline
    &7       &300        &0          &0          &0.83   &2D       \\
    strong magnetized&8       &0          &300        &0          &0.83   &2D       \\
    &9       &0          &0          &300        &0.83   &2D       \\
    \hline
    3D &10      &0          &150        &0.0        &3.35   &3D       \\
    medium magnetized&11      &106        &106        &0.0        &3.35   &3D        \\  
    \enddata
\end{deluxetable}

The previous numerical simulation results \citep{mackey2011effects} suggest that magnetic fields are dynamically important in the formation of pillar structures. Sugitani et al. 2007 \citep{sugitani2007near} used near-infrared polarimetry to measure the magnetic field in the Eagle Nebula, and they found a difference in the direction of magnetic fields between the pillars and the surrounding photo-ionized clouds. K. Pattle et al. 2018 \citep{pattle2018first} used the $850 ~\mathrm{\mu m}$ polarized light to give a high-resolution map of the magnetic field in the dense gas of the photo-ionized pillars. They show that the magnetic fields run along the length of the pillars, perpendicular to and decoupled from the field in surrounding clouds, and this configuration can also support the pillars. These observation results show that there exist large-scale ordered magnetic fields near the pillar, but the morphology of the magnetic field in Eagle Nebula is complicated, and the finer magnetic field structure is still unclear.

In this section, we study the effect of different magnetic field components on forming a pillar-like structure. The initial thermal pressure of ionized blow-off gas plasma is about $P_T \approx 3\times 10^{11} ~ \mathrm{dyn/cm^2}$, then we define the initial plasma beta as $\beta_{initial} = P_{T} / P_{B} = P_{T} / (B^2/8\pi)$. \textcolor{black}{The simulation in this study includes cases of magnetic field components perpendicular ($B_x,~B_z$) and parallel ($B_y$) to the pillars, as shown in Table~\ref{table-case}. The cases can be categorized into three types: weakly magnetized (cases 1 to 3), medium magnetized (cases 4 to 6 and cases 10 to 11), and strongly magnetized (cases 7 to 9). Cases 10 to 11 correspond to 3D simulations. The magnetic field is frozen in the plasma flux due to the large magnetic Reynolds number \citep{crutcher2012magnetic}.}

\begin{figure}[htb]
    \centering
    \includegraphics[width=16cm]{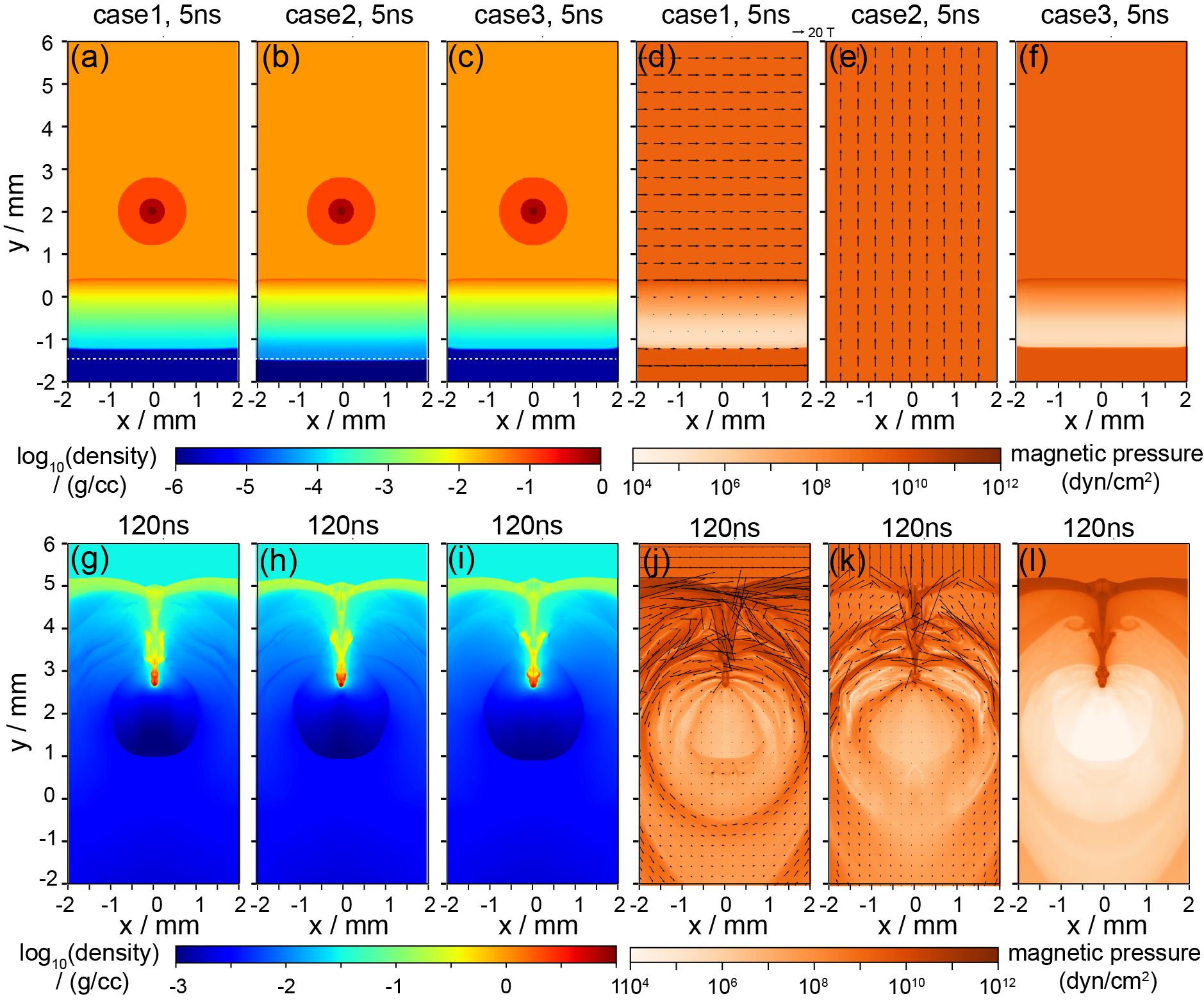} 
    \caption{{\bf \textcolor{black}{ Weakly magnetized cases.}} Panels (a) to (c) correspond to the logarithmic mass density in $\rm g/cm^3$ for case 1, case 2, and case 3 at 5 ns respectively, and the color bar is given at the bottom of these panels. The white dot line shows the front position of the blow-off gas plasma for the case without magnetic fields (as shown in \ref{radiation-material}(b)). Panels (d) to (f) are the corresponding magnetic pressure for these cases at 5 ns \textcolor{black}{in logarithmic values}, and the black arrows represent the magnetic field lines. The distributions of mass density at 120 ns for these cases are shown in panel (g) to (i), and the corresponding magnetic pressures are given in (j) to (l) \textcolor{black}{in logarithmic values}.}
    \label{weak-mag}
\end{figure}

{\bf Weakly magnetized cases} 

The weakly magnetized cases of different magnetic components are shown in figure \ref{weak-mag}. The white dot line in \ref{weak-mag}(a) to \ref{weak-mag}(c) represents the front position of the blow-off gas plasma for the case without magnetic fields. We find that the magnetic field components (case 1, case 3) perpendicular to the pillar will suppress the movement of the blow-off low-density plasma. At the same time, the magnetic field is also significantly compressed and amplified in these cases. The peak strength of the magnetic field is $60 \times 10^4 ~\mathrm{Gs}$, which is $3$ times higher than the initial magnetic field. However, the magnetic pressure ($1\times10^{10} ~\mathrm{dyn/cm^2}$) is still much smaller than the thermal pressure value, so the hydrodynamic effect is still dominant. For case 2 (\ref{weak-mag}(b)), the front position of the blow-off gas plasma is similar to the case without magnetic fields, and there is no significant change in the distribution of the magnetic field in the early stage (\ref{weak-mag}(e)). 

At $120~\mathrm{ns}$, the morphologies of pillar-like structure (for case 1 to case 3) are almost the same as the case without magnetic fields (\ref{pillar-form}(g)), but we can see that the field orientation is significantly changed by the dynamics of the photon-ionization process. The magnetic Alfv\'en velocity within the unshocked material can be calculated as $v_A = B/\sqrt{4\pi \rho} = 20\times10^4 ~\mathrm{Gs} / \sqrt{4\pi \times 0.03~ \mathrm{g/cm^3}} \approx 5.3\times 10^5 ~\mathrm{cm/s}$, which is much smaller than the shock speed. Then, the formation of a pillar-like structure is much faster than the time of the magnetic field relaxing to a lower energy configuration. For case 1 (\ref{weak-mag}(j)), the magnetic fields roughly keep the initial orientation near the shock front, while within the pillar-like structure, we find that the magnetic field is compressed and amplified to $100\times 10^4 ~\mathrm{Gs}$, and the direction of the magnetic field becomes parallel to the pillar, which is similar to the high-resolution observations \citep{sugitani2007near,pattle2018first}. For case 2, at the position of the shock front in $120~\mathrm{ns}$ (\ref{weak-mag}(k)), a perpendicular component appears. \textcolor{black}{This is because when X-rays interact with a spherical high-density clump, a plasma is generated that expands approximately spherically, leading to a significant change in the direction of the magnetic field in case 2.} Within the pillar-like structure, the magnetic fields still keep parallel to the pillar. For the $B_z$ component case, magnetic field lines cannot be drawn directly in 2D. According to the principle of symmetry, the evolution should be similar to the $B_x$ case, and as shown in Fig.~\ref{weak-mag}(l), the distribution of magnetic pressure is also similar to Fig.~\ref{weak-mag}(j).

{\bf Medium magnetized cases} 

\begin{figure}[htb]
    \centering
    \includegraphics[width=16cm]{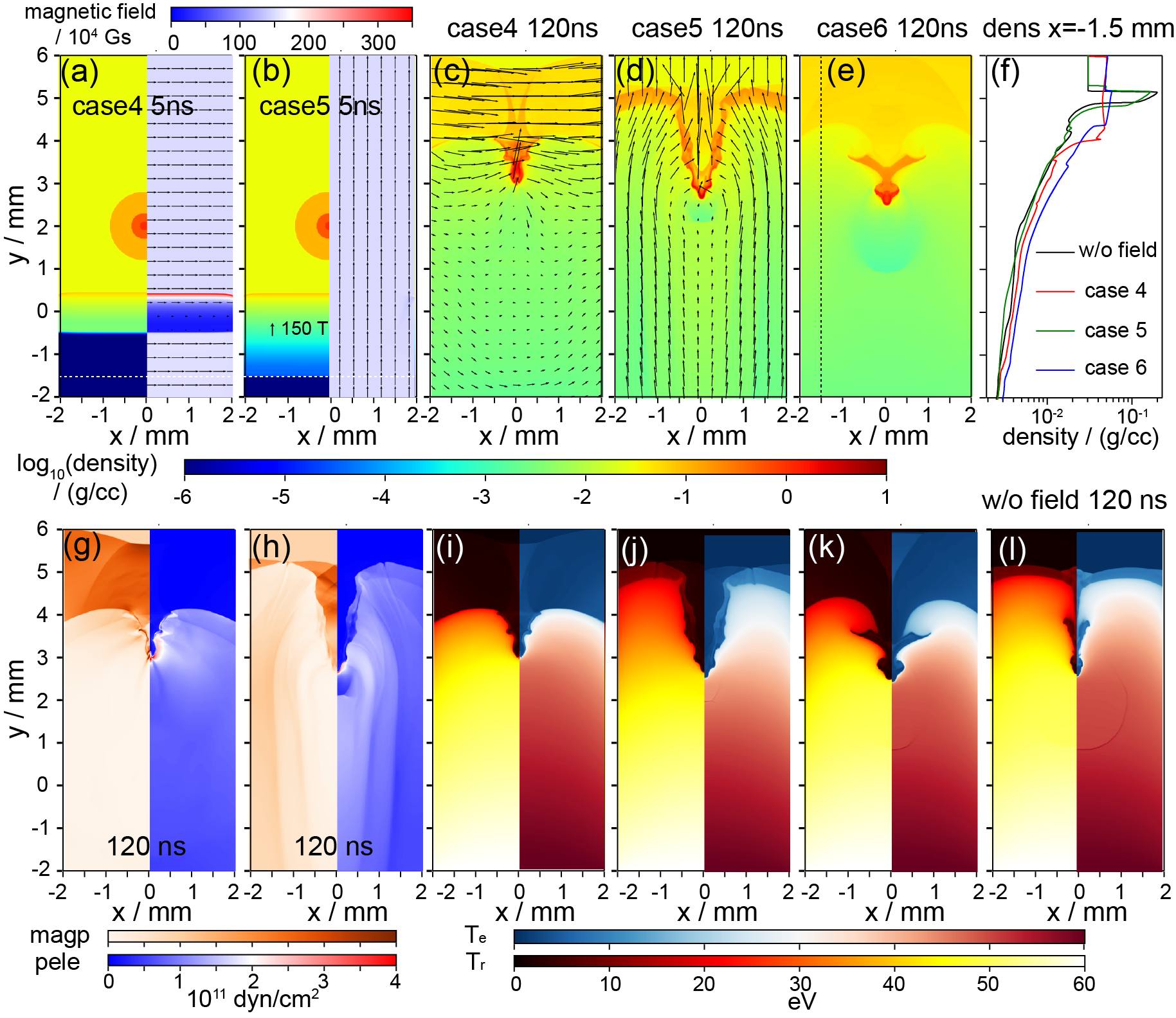} 
    \caption{{\bf \textcolor{black}{ Medium magnetized cases.}} The left sides of panels (a) and (b) correspond to the logarithmic mass density in $\rm g/cm^3$ for case 4, case 5 at $5~\mathrm{ns}$, and the right sides correspond to the distribution of magnetic fields. Panels (c) to (d) plot the logarithmic mass density for cases 4 to 6 at $120~\mathrm{ns}$. The mass densities of low-density foam at $x=-1.5~\mathrm{mm}$ for different cases are shown in panel (f). The left sides of panels (g) and (h) plot the magnetic pressure for case 4, and case 5 at $120~\mathrm{ns}$, and the right sides are the corresponding thermal pressure. The left sides of panels (i) to (l) are the radiation temperature in eV for case 4 to case 6 and case without magnetic fields, and the right sides of these panels are the corresponding electron temperature.}
    \label{medium-mag}
\end{figure}

The medium magnetized cases of different magnetic components are shown in figure \ref{medium-mag}. The magnetic components perpendicular to the pillar (case 4,6) can suppress the expansion of low-density plasma more effectively than the weakly magnetized cases. As shown in \ref{medium-mag}(a), the $B_x$ component effectively suppresses the expansion of low-density plasma, and the magnetic field is compressed and amplified to 350T (on the right-side of \ref{medium-mag}(a)). The amplified magnetic pressure is already close to the thermal pressure, which plays an important role in the subsequent evolution of plasma dynamics. For the initial components parallel to the pillar (case 5), low-density plasma motion along the magnetic field lines is barely affected and there is also little change in the magnetic field distribution in the early stage (\ref{medium-mag}(b)). 

Here, we mainly focus on the formation of a pillar-like structure in the late stage. The distributions of logarithmic mass density at $120 ~\mathrm{ns}$ for different cases are plotted in \ref{medium-mag}(c) to \ref{medium-mag}(e), and the corresponding magnetic field lines are given in \ref{medium-mag}(c) and \ref{medium-mag}(d) through black arrow lines. For case 4 (\ref{medium-mag}(c)), a very narrow pillar-like structure forms. The magnetic field inside the cloud is compressed and amplified, so the magnetic pressure can provide the balance with the ablation pressure, and then the shock wave propagation speed drops than without the magnetic field case. We also find a noticeable perturbation in the undisturbed low-density foam behind the shock front, and the perturbation thickness is about $2~\mathrm{mm}$. This perturbation may be due to Alfv\'en waves, which are perpendicular to the magnetic field. The mean shock velocity is about $4- 5 \times 10^{6} ~\mathrm{cm/s}$, and the magnetic Alfv\'en velocity (after magnetic field compression and amplification) is about $9 \times 10^{6} ~\mathrm{cm/s}$, so the disturbance of Alfv\'en waves can transfer faster. The initial magnetic field lines are perpendicular to the pillar and at $120 ~\mathrm{ns}$, most of the field lines still remain perpendicular. This indicates that with the increase of the magnetic field, especially when the magnetic pressure and thermal pressure are comparable, the dynamic behavior of the plasma is difficult to significantly change the topology of the magnetic field.

For case 5, a long and stable pillar is formed, and the morphology of the pillars is most similar to the POC. As shown in Fig.~\ref{medium-mag}(d), the field lines are perpendicular to the boundary of the pillar, and within the pillar, the magnetic field still remains in the initial direction, but the strength has been amplified to $300~ \mathrm{T}$. The amplified magnetic pressure prevents the shock waves on both sides of the high-density clump from converging toward the axis (as shown in the right side of \ref{medium-mag}(h)), and the magnetic field strength is large enough to magnetically support the sides of pillars against collapsing radially, so a long and stable pillar can be formed. For the $B_z$ component case, a triangle-like structure is formed at the head of the pillar, and a vortex appears after the triangle head. This vortex instability may be due to the large Larmor radius instability within the vertical magnetic field\citep{ripin1987large}. This instability is mainly driven by gravitation or acceleration drifts such that $v_g \gg v_{di}$, where $v_g = g/\omega_{ci}$, $v_{di} = T_i / ZeL_n B$ are the gravity drift and diamagnetic drift respectively, and $g,\omega_{ci}, T_i, Z, e, L_n, B$ are effective gravity, ion gyro-frequency, ion temperature, ion charge state, the elementary charge, plasma-density-gradient scale length, and strength of the magnetic field. Here, the condition of $v_g \approx 4700 \gg v_{di} \approx 5$ is met, so the large Larmor radius instability can grow up at the boundary of pillar, and it is similar to the experiment results \citep{tang2020observation}. The mass densities of low-density foam at $x=-1.5~\mathrm{mm}$ for different cases are shown in Fig.~\ref{medium-mag}(f). We find that the mass density distribution of case 4 and case 6 are similar. The peak number density of case 5 is similar to the case without fields, and it is about $3- 4$ times higher than the peak density in case 4 or 6. 

\begin{figure}[htb]
    \centering
    \includegraphics[width=16cm]{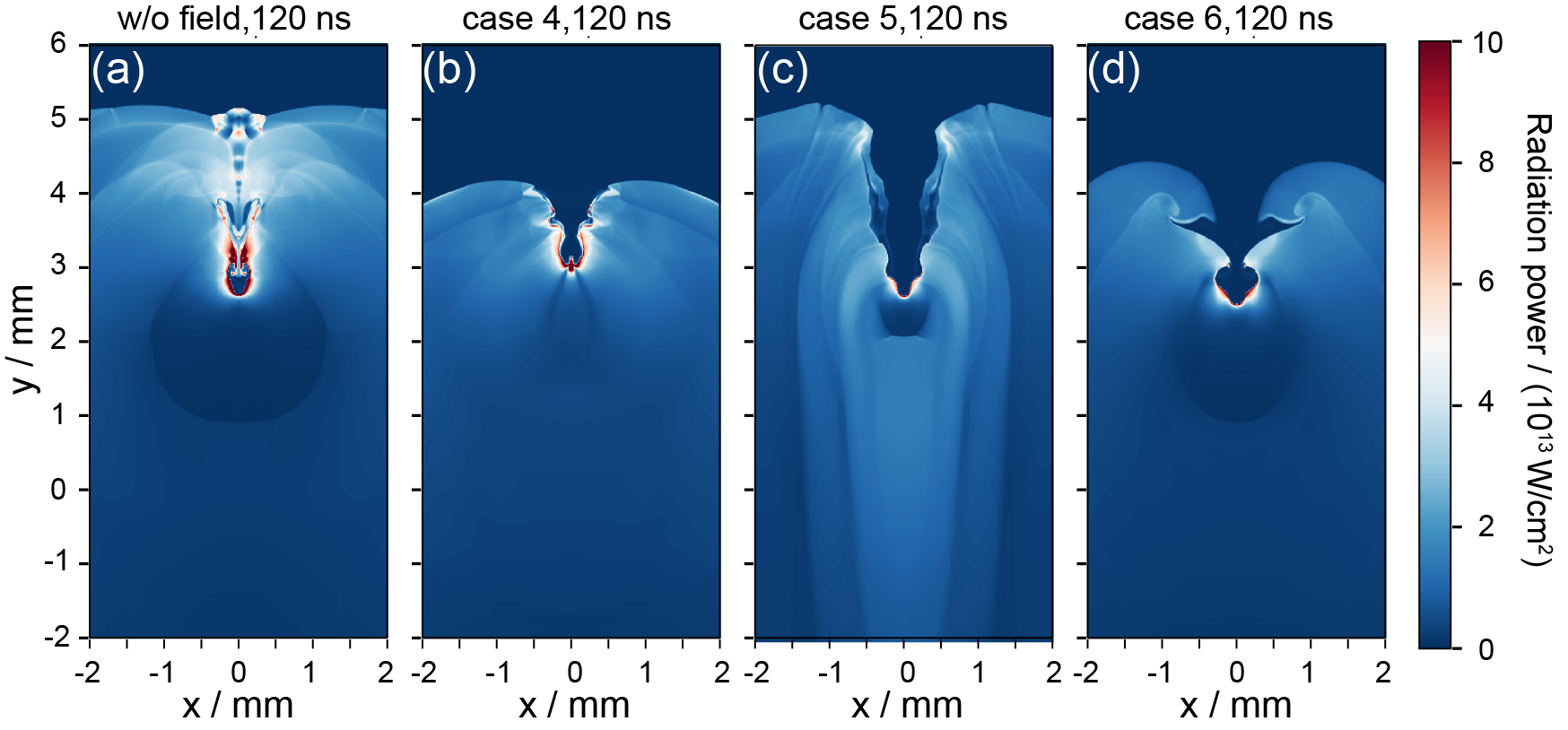} 
    \caption{{\bf \textcolor{black}{ Radiation for medium magnetized cases.}} The distributions of radiation power for different cases at $120~\mathrm{ns}$ are shown in panels (a) to (d).}
    \label{medium-mag-radiation}
\end{figure}

The distributions of electron temperature and radiation temperature for cases 4 to 6 at $120~\mathrm{ns}$ are shown in Figs.~\ref{medium-mag}(i) to \ref{medium-mag}(k), and \ref{medium-mag}(l) plots the case without magnetic fields. We find that in the cases with magnetic fields, the head, middle, and tail of the pillar maintain a low-temperature structure, and the radiation temperature inside the pillar is also very low, so the tail will not be separated from the head. In Fig.~\ref{medium-mag}(l), for the case without magnetic fields, only the head of the pillar keeps a low temperature, while the tail has been effectively heated, whose temperature is about $30- 40 ~\mathrm{eV}$.

The gas emission in recombination radiation (e.g. $H_\alpha$ or X-ray) is one of the most important observational features for the POC \citep{linsky2007chandra,sofue2020co}, and the most luminescent part is located on the pillar's head, followed by a long shadow tail (as shown in Fig.\ref{setup}(b)). In the experiments, the bremsstrahlung emission, as we mentioned above, is the main radiation mechanism for the high-temperature low-Z plasma \citep{nicolai2006plasma}, and the radiation maps for medium magnetized cases are shown in figure \ref{medium-mag-radiation}. \ref{medium-mag-radiation}(a) corresponds to the case without magnetic fields, and the distinct radiation features appear on both the head and tail of the pillar, which is quite different from the observation results. For cases with magnetic fields, their radiation signature is very similar to the observations, where the most intense radiation is located on the pillar's head, followed by a long shadow tail. The case in which the $B_y$ component dominates (case 5, where the magnetic field lines are parallel to the pillar) is most similar to the observed result.

{\bf Strong magnetized cases}

\begin{figure}[htb]
    \centering
    \includegraphics[width=16cm]{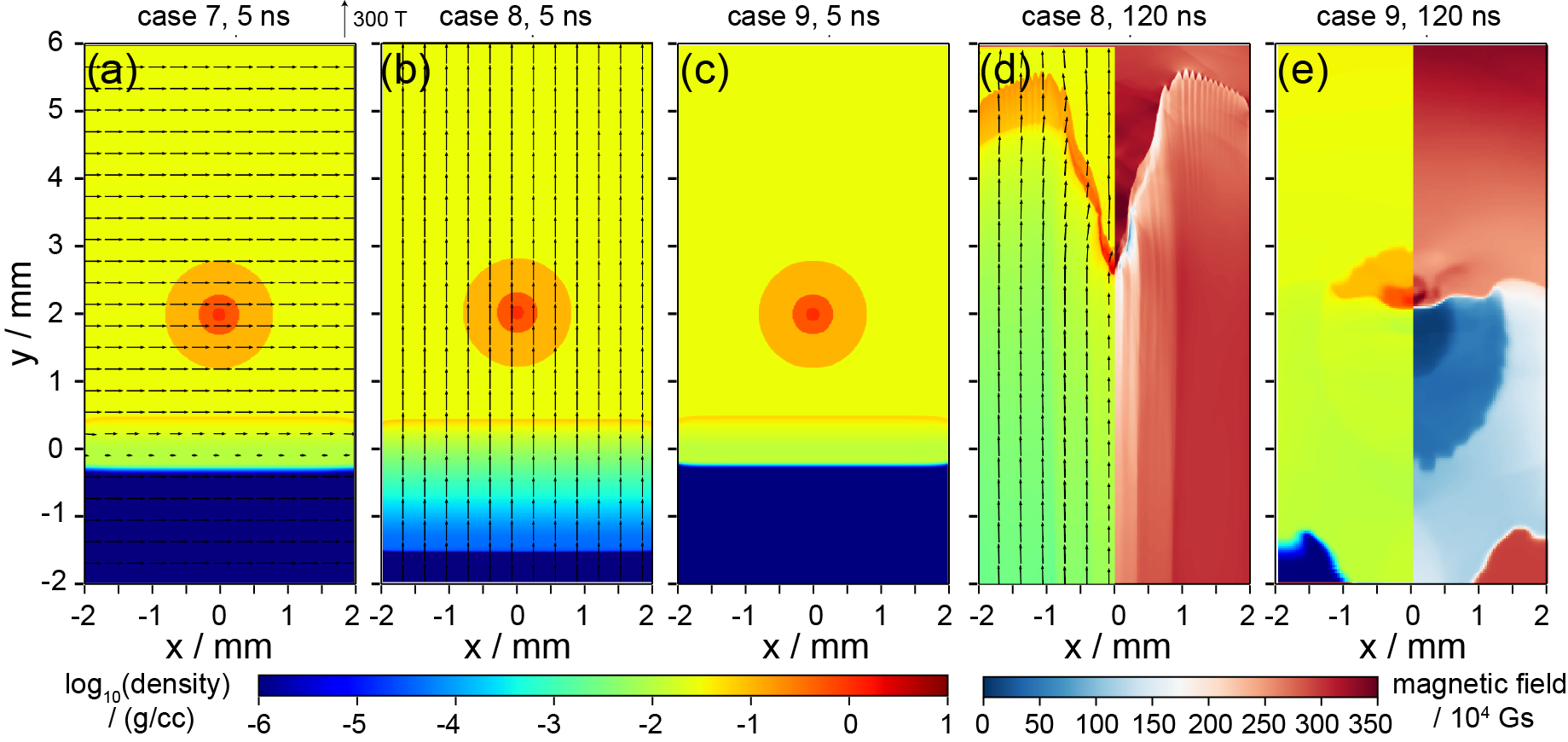} 
    \caption{{\bf \textcolor{black}{ Strong magnetized cases.}} Panels (a) to (c) correspond to the \textcolor{black}{logarithmic mass density in $\rm g/cm^3$ for case 7 to case 9 at $5~\mathrm{ns}$.} The left sides of panels (d) and (e) are the distribution of mass density at $120~\mathrm{ns}$, and the right side is the corresponding magnetic fields.}
    \label{strong-mag}
\end{figure}

If we continue to increase the magnetic field strength, when the initial magnetic pressure is larger than the thermal pressure, the plasma dynamics are significantly affected by the magnetic field. As shown in Fig.~\ref{strong-mag}(a) and (c), the movement of blow-off gas flows is significantly suppressed when the magnetic field component is perpendicular to the pillar, and almost no obvious forward propagating shock is formed. The initial magnetic field is compressed and amplified to about 350 T. Compared with the weak magnetized cases and medium magnetized cases, the magnetic field is amplified to a smaller extent, only 1.1 times the initial field strength. \textcolor{black}{And the magnetic pressure inside the pillar is also much higher, but this does not cause rapid lateral diffusion and destruction of the pillar structure. For strong magnetized cases, the topology and distribution of the magnetic field are not significantly affected because the thermal pressure of the blow-off gas is lower than the initial magnetic pressure. And total pressure of residual magnetic pressure outside the pillar and the thermal pressure can maintain the stability of the pillar structure.} As shown in Fig.~\ref{strong-mag}(e), pillar-like structures cannot be formed even at later stages (120~ns), and only an arc-like structure is formed. For case 9 (\ref{strong-mag}(b)), since the plasma is not hindered along the direction of the magnetic field lines, even if the initial plasma $\beta$ is less than 1, the motion of the plasma is not significantly different from the case without magnetic fields. Similar to the medium case, the magnetic pressure can support the sides of pillars against collapsing radially under pressure from the surrounding hot ablated plasma (\ref{strong-mag}(d)), but the head of the pillar forms a more pointed structure.

In brief, for the weakly magnetized cases, the plasma thermal pressure clearly dominates and the morphology of pillar-like structure is similar to the case without magnetic fields. The magnetic field orientation is obviously changed by the plasma dynamics, and we find the magnetic field lines all become parallel to the pillar within the pillar structure in a later stage for different cases. \textcolor{black}{In medium magnetized cases, the initial magnetic fields are compressed and amplified from $150~\mathrm{T}$ to $300~ \mathrm{T}$, resulting in an increase in magnetic pressure by $4-5$ times. This magnetic pressure is close to the thermal pressure. And it is sufficient to support the sides of pillars against radial collapse under pressure from the surrounding hot plasma, thereby maintaining the structure of the pillar and increasing its longevity.} According to the radiation maps of medium magnetized cases, when the magnetic field lines are parallel to the pillar, the radiation feature is most similar to the observed results, and the field orientation within the pillar is also consistent with the observation results \citep{sugitani2007near,pattle2018first}. For the strong magnetized cases, the formation of pillars is significantly suppressed when the magnetic field component ($B_x, B_z$) is parallel to the shock front. 

\subsection{3D simulation results}

\begin{figure}[htb]
    \centering
    \includegraphics[width=16cm]{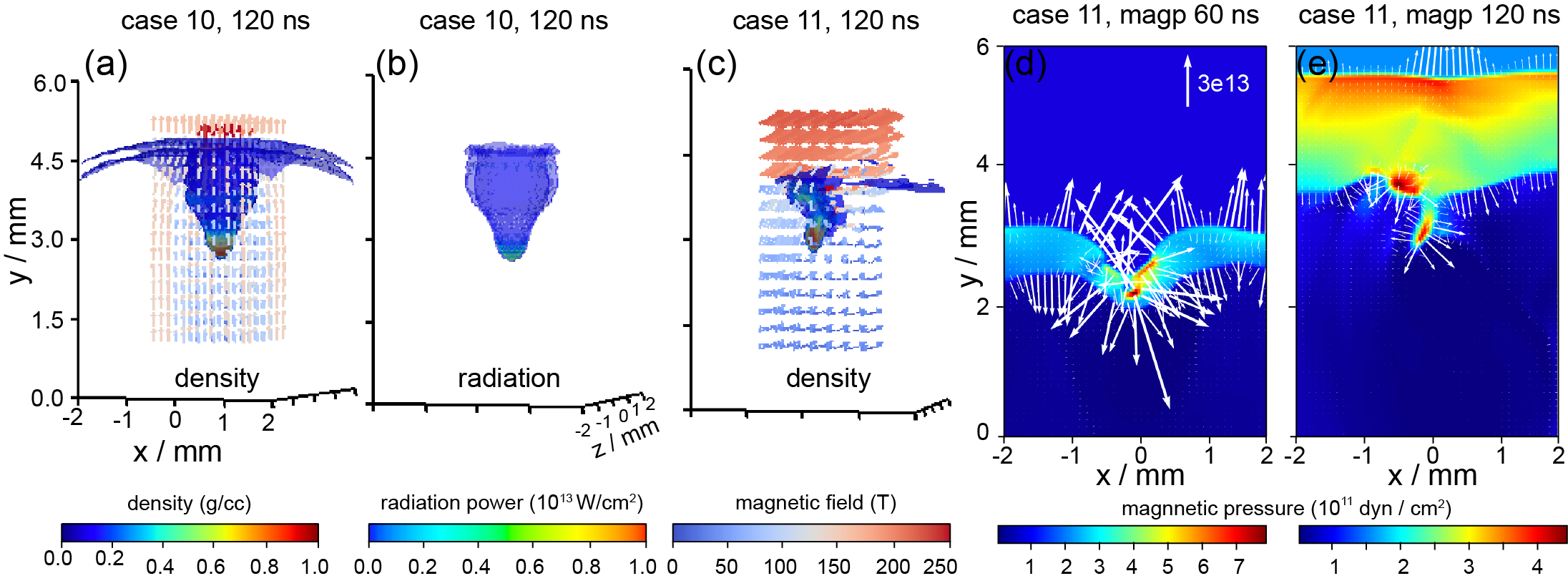} 
    \caption{{\bf \textcolor{black}{3D simulation results.}} Panels (a) and (b) correspond to the distribution of mass density in $\rm g/cm^3$ and radiation power for case 10 at $120~\mathrm{ns}$. Panel (c) is the distribution of mass density for case 11, and the color bar for mass density is given under panel (a). The arrows in panels (a) and (c) represent the magnetic field vector and the colors on the arrows represent the strength (the color bar is given under panel (c)). \textcolor{black}{Panel (d) and (e) plot the distribution of magnetic pressure for case 11 at $60~\mathrm{ns}$ and $120~\mathrm{ns}$ respectively in the XY plane, and the white arrows represent the Lorentz force in the unit of $\mathrm{dyn/cm^{3}}$.}}
    \label{3D-simu}
\end{figure}

In order to avoid dimensional effects due to 2D simulation, we also performed full 3D simulations. As shown in Fig.~\ref{3D-simu}(a) and (b), the setup of case 10 (3D simulation) is the same as case 5 (2D simulation). The distribution of mass density (\ref{3D-simu}(a)) shows a nearly axisymmetric and long pillar structure is formed, and the 2D simulation result (\ref{medium-mag}(d)) is also similar to this result, which confirms the reliability of the 2D simulation. The magnetic field strength before the pillar is reduced (as shown by arrow lines in \ref{3D-simu}(a)), and the magnetic field within the pillar is compressed and amplified to $280~\mathrm{T}$. The magnetic field lines are perpendicular to the pillar boundary outside the pillar. The distribution of radiation shows the most luminescent is located on the head of this pillar structure, and it is also a pillar-like structure. 

However, it is difficult for the magnetic field to be perfectly parallel or perpendicular to the pillar structure in Eagle Nebula. So we refer to Ryutov's model \citep{ryutov2005two}, making a natural assumption that the parallel ($B_y$) and perpendicular ($B_x$) components are initially comparable ($B_x = B_y = 106~\mathrm{T}$), and the total initial strength of the magnetic field remains at $150~\mathrm{T}$. As shown in Fig.~\ref{3D-simu}(c), the perpendicular component of the magnetic field inside the foam is compressed by the shock wave, and magnetic fields left in the shocked blow-off gas is mainly the parallel component, which is well consistent with Ryutov's model \citep{ryutov2005two}. But the mass density distribution map shows that the pillar structure is asymmetrical, not a collimated long pillar structure, which gradually bends from the head to the tail, forming an arc-shaped structure. And the morphology of this asymmetrical pillar is similar to the head of `Column \uppercase\expandafter{\romannumeral1}' in the POC (as shown by the green circle in \ref{setup}(b)). The radiation distribution also has a similar asymmetric structure. 

\textcolor{black}{As shown in Fig.~\ref{3D-simu}(d) and \ref{3D-simu}(e), we find that the distribution of magnetic pressure is asymmetric during the formation of the pillar, and the magnetic pressure is larger on the right side of the pillar due to the shock compressed. Then this would lead to an asymmetry of the Lorentz force, as shown in the white arrows in \ref{3D-simu}(d). The right side of the high-density clump is subjected to greater downward resistance, so compared to the fluid on the left side of the pillar, the movement of the fluid on the right side will be more hindered, then the pillar eventually develops into an asymmetric structure. And at $\rm 120~ns$, We can clearly see that the asymmetrical structure of the magnetic pressure still exists, and this asymmetrical force leads to the generation of an asymmetrical pillar structure in case 11.}

\section{discussion and conclusion}
\label{sec:discussion}
The model of a quasi-homogeneous magnetic field is proposed based on the assumption that there is a large-scale magnetic field in the cloud, and the large-scale magnetic field is also considered to play an important role in the process of star formation \citep{blandford1982hydromagnetic}. In this work, we consider the effect of magnetic fields on the formation of laser-driven scaled pillars of creation. \textcolor{black}{A low-density foam with preexisting triple-density spherical dense clumps (maximum $1 ~ \mathrm{g/cm^{3}}$) are used in our simulations, and such target configuration makes manufacturing in experiments more convenient. We also use our radiation simulation to reproduce the POC experiments in the OMEGA EP facility, and the simulation results are very similar to their experimental results, which show the radiation hydrodynamic simulation is reliable in predicting experimental results. Regarding the generation of the magnetic field in the experiment, we plan to use the pulse-power system \citep{hu2022upgraded} for weakly magnetized cases. For medium and strong magnetized cases, we plan to generate magnetic fields using the laser-driven capacitor coil target method. And in recent years, a large number of magnetic field generation experiments \citep{zhang2018generation} have been conducted on the SG-II laser facility.}

\textcolor{black}{The astrophysical POCs occur on spatial and temporal scales that are typically 10–20 orders of magnitude greater than those of laboratory experiments that are designed to simulate them. As a result, ensuring similarity between the astrophysical phenomenon and its laboratory counterpart becomes a critical issue. According to previous works \citep{ryutov1999similarity,ryutov2001magnetohydrodynamic,ryutov2002scaling}, if the key dimensionless numbers are kept much larger than unity in two systems, such as magnetic Reynolds number, Reynolds number, and P\'eclet number, both systems will behave as ideal compressible hydrodynamic fluids. And the control equations will remain invariant in the two systems when the following transformation conditions are satisfied, $r_{lab} = ar_{ast}$, $ \rho_{lab} = b\rho_{ast}$, $ P_{lab} = cP_{ast}$, $ t_{lab} = a\sqrt{b/c} t_{ast}$, $ v_{lab} = \sqrt{c/b} v_{ast}$, $ B_{lab} = \sqrt{c} B_{ast}$, where a, b, and c are arbitrary positive numbers, and they also called free transformation parameters. The subscripts ``ast" and ``lab" represent the astrophysical pillars and those of the laboratory.}

\textcolor{black}{For the astrophysical POC, the estimated magnetic diffusion time is about $10^{13}$ years \citep{kane2015dynamics}, so the magnetic Reynolds number is much larger than unity (note that the age of pillars is about $10^4 - 10^5$ years).  Viscosity and thermal conductivity are negligible, because of the very large spatial scales involved, so that the Reynolds number and Péclet number are also much larger than unity. In the experiments, these key dimensionless parameters can be calculated as follows.} The Reynolds number is $R_e = \boldsymbol{v} L/ \nu $, where $\boldsymbol{v}, L, \nu$ are the velocity, system size, and viscosity respectively, and the viscosity can be calculated as $\nu = \left(k_{B} T\right)^{5 / 2} / [\pi n_{i} m_{i}^{1 / 2} Z^{4} e^{4} \ln \Lambda]$ \citep{drake2018high}, where $k_{B},~ T,~ n_i,~ m_i,~ Z,~\ln \Lambda$ are Boltzmann constant, temperature, ion number density, ion mass, average ionization and Coulomb logarithm; the magnetic Reynolds number of $R_m = \boldsymbol{v}L / \eta$, where $\eta = c^2 m_e^{1/2} Z e^2 ln\Lambda / [4(k_B T)^{3/2}]$ is the magnetic diffusivity \citep{brandenburg2005astrophysical}; and the Peclet number $P_e = vL / \kappa$, where $\kappa = \left(k_{B} T\right)^{5 / 2} / [m_{e}^{1 / 2} n_{i} Z(Z+1) e^{4} \ln \Lambda]$ is the thermal diffusivity \citep{brandenburg2005astrophysical}. As shown in table \ref{table-scaling}, the Magnetic Reynolds number, Reynolds number, and Peclet number all satisfy the condition of being far greater than the unity, so the scaling laws are valid, and as mentioned above, the radiation cooling parameter is also closer to the astrophysical system. \textcolor{black}{Then, the same set of equations can accurately describe both the astrophysical and laboratory scenarios.} Table ~\ref{table-scaling} gives the detailed comparison between the scaled-up laboratory system and specifically the POC in Eagle Nebula, where the transformation parameters $a = 3.3 \times 10^{-20}, b= 3\times 10^{18}, c= 2.0 \times 10^{18}$ are taken. \textcolor{black}{According to astrophysical observation results \citep{pattle2018first}, the typical magnetic field strength within the pillar is about $100-300~ \rm \mu Gs$, and the directions of the magnetic field are parallel to the cylindrical structure. The range of magnetic field strength corresponding to the laboratory system is approximately 100-300 T after conversion. In this work, this is very similar to the case 5 situation, and our laboratory radiation simulation also indirectly indicates that the pillar structure is mostly collimated and stable when there is a parallel magnetic field component, which corresponds to astronomical observations}. The scaling laws confirm that our laboratory simulation results can be applied to explore the formation mechanism of POC.

\begin{deluxetable}{ccccc}

\tablecaption{Comparison of the Eagle Nebula with the laboratory system under the scaling laws}
\tablenum{2}
\label{table-scaling}

\tablehead{\colhead{Parameters} & \colhead{Equation} & \colhead{Astrophysical system} & \colhead{Scaled-up POC} & \colhead{Laboratory system} } 

\startdata
Length                   & $L$                                & $\sim 5~\mathrm{pc}$                     & $\sim 4.9~\mathrm{pc}$                           & $\sim 5 ~\mathrm{mm}$           \\
Density                  & $n_e$                              & $10^{-20} - 10^{-19} ~\mathrm{g/cc}$     & $\sim 10^{-20} ~ \mathrm{g/cc}$                  & $0.03 ~\mathrm{g/cc}$           \\
Pressure                 & $P$                                & $10^{-9} - 10^{-8} ~ \mathrm{dyn/cm^2}$  & $\sim 5\times 10^{-9} ~ \mathrm{dyn/cm^2}$       & $10^{10} ~ \mathrm{dyn/cm^2}$   \\
Velocity                 & $\boldsymbol{v}$                   & $\sim 10^6 \mathrm{cm/s}$                & $\sim 1.2\times 10^6 \mathrm{cm/s}$              & $10^6 - 10^7 \mathrm{cm/s}$     \\
Timescale                & $t$                                & $10^4 - 10^5 ~ \mathrm{years}$           & $\sim 9.4\times10^4 ~ \mathrm{years}$            & $\sim 120 ~ \mathrm{ns}$        \\
Magnetic field           & $B$                                & $10 - 200 ~\mathrm{\mu Gs}$              & $\sim 106 ~\mathrm{\mu Gs}$                      & $1.5\times 10^5 ~\mathrm{Gs}$   \\
Temperature              & $T_e$                              & $1 ~\mathrm{eV}$                         & \nodata                                          & $100 ~\mathrm{eV}$              \\
Magnetic Reynolds number & $R_m = \boldsymbol{v}L / \eta $    & $>10^{8}$                                & \nodata                                          & $>10^2$                         \\
Reynolds number          & $R_e = vL/ \nu$                    & $>10^7$                                  & \nodata                                          & $>10^7$                         \\
Peclet number            & $P_e = vL / \kappa $               & $>10^7$                                  & \nodata                                          & $>10^4$                         \\
Radiative ratio          & $\chi = \tau_{rad} / \tau_{hd} $   & $10^{-3} \sim 10^{-2}$                   & \nodata                                          & $\sim 10^{-2}$                  \\
\enddata

\end{deluxetable}

In summary, we propose a new experimental scheme to study the formation and support mechanism of POC. Three materials (CH, Al, Au) are used in our simulations, and the Al-doped CH foam should be used as the first choice for experimental targets according to our simulation results, whose radiative ratio is closer to the POC and a strong shock can be launched by the ablation process. And the 2D and 3D RMHD simulations all show that the magnetic fields, especially the $B_y$ component, play an important role in the formation of POC. For the medium magnetized cases, the magnetic pressure can effectively support the sides of pillars against collapsing, and a long collimated pillar structure is formed. The radiation maps are also similar to the observed results. The bending at the head of `Column \uppercase\expandafter{\romannumeral1}' in POC may be due to the non-parallel magnetic field lines according to our 3D RMHD simulations. The similarities between laboratory and astrophysical systems suggest that our results can be applied to explore the formation mechanism of pillars of creation. In addition, the mechanism of magnetic topology changes within the pillars \citep{ryutov2005two} and the multiple pre-existing disturbances \citep{lim20033d,pound2007pillars} are also the key factors in the formation of POC, and we will plan to carry out further related research in the future.

\par

This work is supported by the National Key R\&D Program of China, Grants No. 2022YFA1603200 and No. 2022YFA1603204; Science Challenge Project, No. TZ2018005; National Natural Science Foundation of China, Grant No. 11825502 and 11921006; the Strategic Priority Research Program of Chinese Academy of Sciences Grant No. XDA25050900. L. F. W. is supported by the National Natural Science Foundation of China (Grant No. 11975053). The simulations are carried out on the supercomputers in China.

\bibliography{sample631}{}
\bibliographystyle{aasjournal}

\end{document}